\documentstyle[12pt]{article}
\begin{document}
\vspace{0.5in}
\begin{center}
{\Large \bf The order
$O(\overline{\alpha}~\overline{\alpha}_s)$ and
$O(\overline{\alpha}^2)$ corrections to the decay width of the neutral
Higgs boson to the $\overline{b}b$ pair}

\vspace{0.5in}

{\large \bf A.L.~Kataev}
\vspace{0.1in}

{\it Institute for Nuclear
Research of the Academy of Sciences of Russia,\\
117312 Moscow, Russia}

\end{center}
\vspace{0.1in}
\begin{center}
{\bf Abstract}
\end{center}
We present the analytical expressions for the contributions
of the order
$O(\overline{\alpha}~\overline{\alpha}_s)$ and
$O(\overline{\alpha}^2)$ corrections to the decay width of
the Standard Model Higgs boson into
the $\overline{b}b$-pair.
The numerical value of the mixed QED and QCD correction of order
$O(\overline{\alpha}~\overline{\alpha}_s)$ is comparable with
the previously calculated terms in the perturbative series for
$\Gamma(H^0\rightarrow\overline{b}b)$.

\vspace{2cm}

To be published in Pisma Zh. Eksp. Teor. Fiz. v 66, N5 (1997)\\
(JETP Lett. v 66 (1997)).


\newpage

Amongst the most important problems of the modern high energy physics
are the investigations of the properties of the experimentally
still unobserved Higgs boson of the Standard Model of electroweak
interactions (see e.g. the reviews of Ref. \cite{Rev}).
The analysis of the experimental data of the LEP1 collider resulted
in the derivation of the lower bound on the Higgs boson mass
$M_H>65~GeV$. The experimental program of the LEP2 and LHC
accelerators are thus including the searches of the Higgs boson mass
in the mass region
$65~GeV<M_H\leq 2M_W\approx 160~GeV$, where the main decay mode
of the $H^0$-boson should be the decay to the $\overline{b}b$-pair.

Various effects  of the perturbative QCD corrections to
$\Gamma(H^0\rightarrow\overline{b}b)$ were already calculated
and analyzed by the theoreticians
(see e.g. \cite{BL}-\cite{ChS}).
In Ref. \cite{EW} the leading order electroweak contributions to
this fundamental quantity were also found. The level of the achieved
accuracy of the calculations of $\Gamma(H^0\rightarrow\overline{b}b)$
is putting on the agenda the consideration of the new theoretical
contributions, which
{\it a~priori} can be comparable to the already calculated terms
in the perturbative series for
$\Gamma(H^0\rightarrow
\overline{b}b)$.

In this note, following the analogous calculations of the
QED and QCD corrections to the hadronic
$Z^0$-boson decay width \cite{KatQED},
we are presenting the analytical results for the coefficient of
the order
$O(\overline{\alpha}~\overline{\alpha}_s)$-term in the expression
for  $\Gamma(H^0\rightarrow
\overline{b}b)$ (previously found in the process of the work
of Ref.\cite{KK1}, but still unpublished)
and for the order $O(\overline{\alpha}^2)$-correction to the same
quantity. Our results will be related to the
$\overline{MS}$-scheme.

In the limit  $M_H>>2m_b$ we are interested in (where  $m_b$ is
the  $b$-quark pole mass)
the corresponding perturbative approximation for
$\Gamma(H^0\rightarrow\overline{b}b)$ can be presented in the
following form
$$
\Gamma(H^0\rightarrow\overline{b}b) =
\overline{\Gamma}_0^{(b)}\bigg[
1+\Delta\Gamma_1\overline{a}_s+\Delta\Gamma_2\overline{a}_s^2
+\Delta\Gamma_3\overline{a}_s^3+...
$$
$$
-6\frac{\overline{m}_b^2}{M_H^2}\bigg(1+\Delta\Gamma_1^{(m)}\overline{a}_s
+\Delta\Gamma_2^{(m)}\overline{a}_s^2+...\bigg)+\Delta_t+\Delta^{QED}\bigg]
\eqno (1)
$$
where
$\overline{\Gamma}_0^{(b)}=3\sqrt{2}/(8\pi)G_FM_H\overline{m}_b^2$
and $\overline{m}_b=\overline{m}_b(M_H)$,
$\overline{a}_s=\overline{\alpha}_s(M_H)/\pi$
are the defined in the
$\overline{MS}$-scheme running parameters of QCD, which are
normalized at the Higgs boson pole mass.  The
coefficients $\Delta\Gamma_1$ and $\Delta\Gamma_2$ are well known
\cite{BL},\cite{GKLS}:
$$
\Delta\Gamma_1=\frac{17}{4}C_F\approx 5.667 $$ $$
\Delta\Gamma_2=\bigg[\bigg(\frac{893}{4}-62\zeta(3)\bigg)C_A
-\bigg(65-16\zeta(3)\bigg)Tf+\bigg(\frac{691}{4}-36\zeta(3)\bigg)C_F\bigg]
\frac{C_F}{16}
\eqno(2)
$$
$$
-\pi^2\bigg(\frac{11C_A-4Tf+18C_F}{48}\bigg)C_F\approx 29.147
$$
where $C_F=4/3$, $C_A=3$, $T=1/2$, $f=5$ and $\zeta(3)$=1.202....
The value of the coefficient $\Delta\Gamma_1^{(m)}$
can be extracted from the results of the calculations of Ref.
\cite{ST}:  $$ \Delta\Gamma_1^{(m)}=5C_F\approx 6.667~~~~.
\eqno(3)
$$
The corrections $\Delta\Gamma_2^{(m)}$ and $\Delta\Gamma_3$ were
calculated in the works of Ref. \cite{S} and \cite{ChH}
correspondingly.
We are presenting them in the numerical form in the case of
$f=5$ numbers of active flavours:
$$
\Delta\Gamma_2^{(m)}\approx
14.621~~~,~~~ \Delta\Gamma_3\approx 41.758~~~.
\eqno(4)
$$
The most convenient for phenomenological purposes expressions
for the virtual
$t$-quark contribution in the expression for
$\Gamma(H^0\rightarrow\overline{b}b)$
was recently obtained in the work of Ref.
\cite{ChS}.
It has rather complicated form:
$$
\Delta_t=\overline{a}_s^2\bigg(3.111-0.667L_t-\frac{\overline{m}_b^2}{M_H^2}
(-10+4L_t+\frac{4}{3}ln(\overline{m}_b^2/M_H^2)\bigg)
+\overline{a}_s^3\bigg(50.474-8.167L_t-1.278L_t^2\bigg)
$$
$$
+
\overline{a}_s^2\frac{M_H^2}{m_t^2}\bigg(0.241-0.070L_t\bigg)
+X_t\bigg(1-4.913\overline{a}_s+\overline{a}_s^2(-72.117-20.945L_t)\bigg)
+....
\eqno(5)
$$
where $L_t=ln(M_H^2/m_t^2)$, $X_t=G_Fm_t^2/(8\pi^2\sqrt{2})$ and $m_t$
is the  $t$-quark pole mass.

Let us turn to the calculation of the Quantum Electrodynamic part
$\Delta^{QED}$ in Eq.(1), which is defined as:
$$
\Delta^{QED}=\bigg[\Delta\Gamma_{1,QED}-6\frac{\overline{m}_b^2}{M_H^2}
\Delta\Gamma_{1,QED}^{(m)}\bigg]\frac{\overline{\alpha}}{\pi}
+\Delta\Gamma_{2,QED}\bigg(\frac{\overline{\alpha}}{\pi}\bigg)^2
+\Delta\Gamma_{QED\times
QCD}\frac{\overline{\alpha}}{\pi}\frac{\overline{\alpha}_s}{\pi}
\eqno(6)
$$
where $\overline{\alpha}$=$\overline{\alpha}(M_H)$ is the normalized
on the Higgs boson pole mass QED running coupling constant of the
$\overline{MS}$-scheme.
The coefficients $\Delta\Gamma_{1,QED}$, $\Delta\Gamma_{1,QED}^{(m)}$
and $\Delta\Gamma_{2,QED}$ can be found from the analytic formulae
of Eqs.(2),(3) after the following substitutions:
$C_A\rightarrow0$, $C_F\rightarrow Q_b^2$ ¨ $Tf\rightarrow
(3\sum_{j=u}^{b}Q_j^2+N)$, where $N=3$ is the number of leptons
and $Q_j$ are the charges of quarks of the corresponding flavour. As
the result we obtain:
$$
\Delta\Gamma_{1,QED}=\frac{17}{4}Q_b^2\approx 0.472~~~,~~~
\Delta\Gamma_{1,QED}^{(m)}=5Q_b^2\approx 0.556 ~~~ ,
$$
$$
\Delta\Gamma_{2,QED}=\bigg(\frac{691}{64}-\frac{9}{4}\zeta(3)
-\frac{3\pi^2}{8}\bigg)Q_b^4 -
\bigg(\frac{65}{16}-\zeta(3)-\frac{\pi^2}{12}\bigg)Q_b^2
\bigg(3\sum_{j=u}^{b}Q_j^2+3\bigg) \approx -1.455~~~.
\eqno(7)
$$

In order to calculate the value of the coefficient of the
$O(\overline{\alpha}~\overline{\alpha}_s)$ correction in Eq.(6)
it is necessary to make the following changes in the analytic
expression of Eq.(2):  $C_A\rightarrow 0$,
$Tf\rightarrow 0$, $C_F^2\rightarrow 2C_FQ_b^2$, where factor 2 is
the symmetry coefficient and $C_F=4/3$. After these substitutions
we arrive to the result we are interested in:
$$
\Delta\Gamma_{QED\times
QCD}=\bigg(\frac{691}{24}-6\zeta(3)-\pi^2\bigg)Q_b^2\approx 1.301~~~.
\eqno(8)
$$

It is rather instructive to raise the question about the numerical
values of the contributions considered by us in the expression for
$\Gamma(H^0\rightarrow\overline{b}b)$.
Here we will limit ourselves by the consideration of the hypothetical
case $M_H\sim M_Z\approx 91~GeV$, which will allow us to simplify
the related numerical estimates. In this case the value of the
parameter $\overline{\alpha}(M_H)$ will be practically
undistinguished from the high-energy value of the QED invariant
charge $\alpha_{inv}(M_Z)\approx 1/129$, determined in the number
of works on the subject (see e.g. Ref. \cite{alpha}).
Other parameters of the theory will be fixed as : $m_b\approx
4.62~GeV$, $m_t\approx 175~GeV$, $G_F\approx 1.166\times
10^{-5}~GeV^{-2}$, $X_t\approx 3.2\times 10^{-3}$,
$\overline{m}_b(M_Z)\approx 2.8~GeV$ (which corresponds to the
central value of the result of the analysis of the DELPHI
collaboration data for the rate of the heavy-quark production in the
3-jet events \cite{bmass}) and $\alpha_s(M_Z)\approx 0.117$ ( it
corresponds to the central value of the QCD coupling constant,
extracted recently from the next-to-next-to-leading order QCD
analysis of the Tevatron data for the $xF_3$ structure function of
$\nu N$ deep-inelastic scattering \cite{KKPS}).

Substituting the above given input data into Eq.(6) and taking into
account the numerical values of the related coefficients (see Eqs.(7),(8))
we find the following numerical estimates of the different QED
contributions into the factor $\Delta^{QED}$:
$$
\Delta^{QED}=1.16\times
10^{-3}-7.79\times 10^{-6}-8.86\times 10^{-6} +1.19\times
10^{-4}~~~~~.  \eqno(9)
$$

Thus the correction of order
$O(\overline{\alpha}~\overline{\alpha}_s)$
turn out to be the order of magnitude smaller then the calculated
in Ref.\cite{ChH} four-loop QCD correction
$\Delta\Gamma_3\overline{a}_s^3$, which under our assumptions gives
the following numerical contribution into
$\Gamma(H^0\rightarrow\overline{b}b)/\overline{\Gamma}_{0}^{(b)}$:
$+2.16\times10^{-3}$. However, it is important to include
selfconsistently the order
$O(\overline{\alpha}~\overline{\alpha}_s)$-term into the final
expression for
$\Gamma(H^0\rightarrow\overline{b}b)$. Indeed, its numerical value
turns out to be comparable with the previously calculated in
Ref.\cite{S} term of order
$O(\overline{a}_s^2\overline{m}_b^2/M_H^2)$
and with the presented in
Ref.\cite{ChS} corrections of order
$O(\overline{a}_s^2M_H^2/m_t^2)$, $O(X_t\overline{a}_s)$ and
$O(X_t\overline{a}_s^2)$. Indeed, in the considered by us case
these corrections have the following order of magnitude:
$-1.15\times10^{-4}$ and $1.24\times 10^{-4}$, $-5.85\times 10^{-4}$,
$-1.98\times 10^{-4}$.  We hope to include the information about
all these corrections into the computer code
SEEHIGGS (for the discussions of its present possibilities see the
note of Ref. \cite{KK3}).

\vspace{1cm}
{\bf Acknowledgments}
\vspace{0.5cm}

We are grateful to V.T.~Kim for the pleasant collaboration, which was
started in 1992 during our simultaneous stay at CERN.

This work was done within the framework of the scientific program
of the Projects N 96-01-01860; N 96-02-18897, supported by
the Russian Foundation of the Fundamental Research.

\end{document}